\documentclass[%
 aip,
 amsmath,amssymb,
 reprint,%
]{revtex4-2}
\usepackage{graphicx}
\usepackage{dcolumn}
\usepackage{bm}

\usepackage[utf8]{inputenc}
\usepackage[T1]{fontenc}
\usepackage{mathptmx}
\usepackage{etoolbox}

\makeatletter
\def\@email#1#2{%
 \endgroup
 \patchcmd{\titleblock@produce}
  {\frontmatter@RRAPformat}
  {\frontmatter@RRAPformat{\produce@RRAP{*#1\href{mailto:#2}{#2}}}\frontmatter@RRAPformat}
  {}{}
}%
\makeatother
\begin{document}

                                                                                                                                                                                                                                                                                                                                                                                                                                                                                                                                                                                                                                                                                                                                                                                                                                                                                                                                                                                                                                                                                                                                                                                                                                                                                                                                                                                                                                                                                                                                                                                                                                                                                                                                                                                                                                                                                                                                  \title{Thermal hot-carrier breakdown in metasurface structures  based on coplanar arrays of graphene microribbons connected with wide-gap bridges}
                                                                                                                                                                                                                                                                                                                                                                                                                                                                                                                                                                                                                                                                                                                                                                                                                                                                                                                                                                                                                                                                                                                                                                                                                                                                                                                                                                                                                                                                                                                                                                                                                                                                                                                                                                                                                                                                                                                                  \author{V.~Ryzhii$^{1,2*}$, M.~Ryzhii$^{3}$, M. S. Shur$^4,5$, T. Otsuji$^{6,7,8}$, and  C.~Tang$^{1,2}$,
}
\address{
$^1$Research Institute of Electrical Communication,~Tohoku University,~Sendai~ 980-8577,
Japan\\
$^2$Frontier Research Institute for Interdisciplinary Sciences,
Tohoku University, Sendai 980-8578, Japan\\
$^3$School  of Computer Science and Engineering, University of Aizu, Aizu-Wakamatsu 965-8580, Japan\\
$^4$Electronics of the Future, Inc., Vienna, VA 22181-6117,\\
 USA\\
$^5$Department of Electrical,Computer, and Systems Engineering, 
Rensselaer Polytechnic Institute,~Troy,~New York~12180, USA\\
$^6$ International Research Institute of Disaster Science, Tohoku University,
Sendai 980-8578, Japan\\
$^7$Center of Excellence ENSEMBLE3 Ltd., Warsaw 01-919,\\
 Poland\\
$^8$ UNIPRESS Institute of High Pressure Physics of the Polish Academy of Sciences, Warsaw 02-822,\\ Poland\\
*{Author to whom correspondence should be addressed: v-ryzhii@gmail.com}}

\begin{abstract}
We  analyze the thermal and electrical characteristics
of the metasurface consisting of 
 the coplanar interdigital array of the graphene microribbons (GMRs) connected by nanobridges (NBs). These nanobridges could be implemented using graphene nanoribbons (GNRs), single-wall semiconducting carbon nanotubes (CNTs), or black-arsenic-phosphorus (b-AsP) nanostructures. The bias voltage applied between neighboring GMRs indices electron and hole two-dimensional systems in the GMRs and induces thermionic currents flowing through connecting NBs. The resulting self-heating increases thermionic currents providing an effective positive feadback between the carrier effective temperature and the injected currents. This mechanism may lead to thermal breakdown enabling threshold behavior of current-voltage characteristics and resulting in the S-shape of these characteristics. The devices based on the GMR/GNR, GMR/CNT, and GMR/AsP metasurface structures can be used as fast voltage-controlled current switches, sensors,  thermal terahertz and infrared sources, and other devices. 
\end{abstract}

\maketitle\section{Introduction}

\begin{figure*}[t]
{\includegraphics[width=16.0cm]{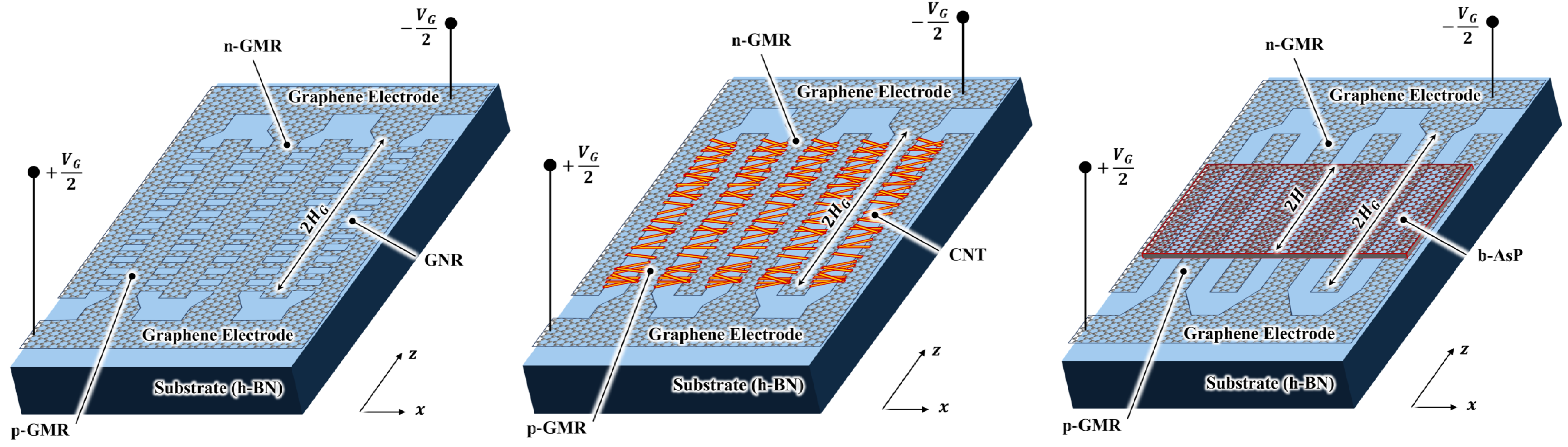}}
\caption{Schematic  view of (a)   GMR/GNR, (b) GMR/CNT, and GMR/AsP metasurface structures
with GL contacting pads.}
\label{fig1}
\end{figure*}

Hybrid heterostructures based on graphene  layers (GLs), in particular, graphene microribbons (GMRs)  connected by nanobridges, such as graphene nanoribbons (GNRs),  single-wall semiconducting carbon nanotubes (CNTs) networks, or the black-arsenic-phosphorus (b-AsP) films, grown on the hexagonal boron nitride (h-BN)
 substrates show great promise  for the development
of different terahertz and infrared devices.~\cite{1,2,3,4,5,6,7,8,9,10,11,12,13,14,15,16,17,18,19,20,21,22,23}
The realization of
this potential requires an appropriate band alignment, the ability to adjust the energy gaps in CNTs and b-AsP by varying the CNT diameters or the number of the atomic layers and the As fraction in b-AsP films, and high-quality interfaces between the GMRs, nanobridges, and h-BN substrate. 

Recently,~\cite{24} we
predicted the effect of thermal hot-carrier breakdown and formation of the S-shaped
current-voltage characteristics in the coplanar interdigital structures consisting of the
GM arrays connected by GNRs.
Such GMR/GNR structures can be formed by perforating pristine uniform graphene layers.  The operation of such devices relies on the carrier
heating  in the GMRs, which enhances the currents through the GNR bridges. A key feature of these devices is the positive feedback
mechanism between the effective carrier temperature and the injected
current, particularly when the current has a thermionic origin.

In this paper, we analyze the GMR/NB structures placed on the h-BN substrate (or similar material) and having longer distances between the GMRs connected by the NBs with a constant width.
In such  structures, NBs form trapezoidal barriers in contrast to previously considered structures with shorter GNRs. We show that these 
proposed structures offer
broader opportunities for  terahertz and infrared sources across various applications, including the fast current switches and incandescent
 terahertz and infrared sources.
The alternating n-type and p-type GMRs are biased via conducting interdigitated  contact pads to create two-dimensional electron and hole systems in the adjacent GMRs.
As a result, the arrays form metasurfaces. 
We show that due to positive feedback effect, the GMR/NB metasurfaces exhibit sharp (threshold) 
voltage dependences of the carrier effective temperature and terminal current-voltage characteristics, which can become many-valued (S-type).  
A detailed
consideration of the thermal and electrical characteristics in a wide ranges of the carrier
effective temperatures and  carrier densities requires a substantial generalization of
the device model  compared with those used previously.

\section{GMR/GNR and GMR/AsP metasurface  structures} 

Figure~1 shows the GMR/GMR, GMR/CNT, and GMR/AsP metasurface structures under consideration placed on a h-BN layer (substrate).
The contact graphene electrodes (pads) are made of metals providing ohmic contacts with the pertinent ends
of the n- and p-GMRs. In particular, these electrodes  can be formed from doped graphene  lines or must be  wide to provide a sufficiently high conductance.
Hence, 
the GMR/NB arrays constitute  systems of interdigitated and connected GMRs of  
 the length and width equal to $2H_G$ and $2L_G$, respectively, whereas  
the NB length (spacing between neighboring  GMRs) 
is $2L$. The width  of the GNRs and b-AsP bridges are $2h$ and $2H$, respectively. The structure under consideration includes $M$ pairs of the GMRs ($M$ can vary from unity to a rather large number). The number of the GNRs (CNTs) between each GMR pair
 can also be different.  Naturally, $H \leq H_G$ and $h \ll H_G $ (with $(2N-1)h < H_G$.
 The GNR width $2h$ and the CNT diameter are chosen to yield proper values of the  energy gap and the respective energy barrier.

The bias voltage $V_G$ applied between the contacting pads induces the n-type
and p-type two-dimensional electron and hole gases. Figure~2 shows the energy band diagram of one device period under the bias voltage $V_G$.
The NBs form the energy barrier $\Delta$ for the electrons in the n-GMRs andfor the holes in the p-GMRs, which, for simplicity, are assumed to be equal.
The operation of the devices based on the metasurface structures under consideration
is associated with the thermionic currents between the GMRs via the NBs, which are determined by the carrier effective temperatures in the GMRs. These electron and hole temperatures depend on the inter-GMR hole and electron currents. The latter can
lead to the positive temperature-current feedback  enabling the many-valued (S-shape)  carrier effective temperature-voltage and  current-voltage characteristics of the metasurface.

\begin{figure}[b]
\centerline{\includegraphics[width=6.0cm]{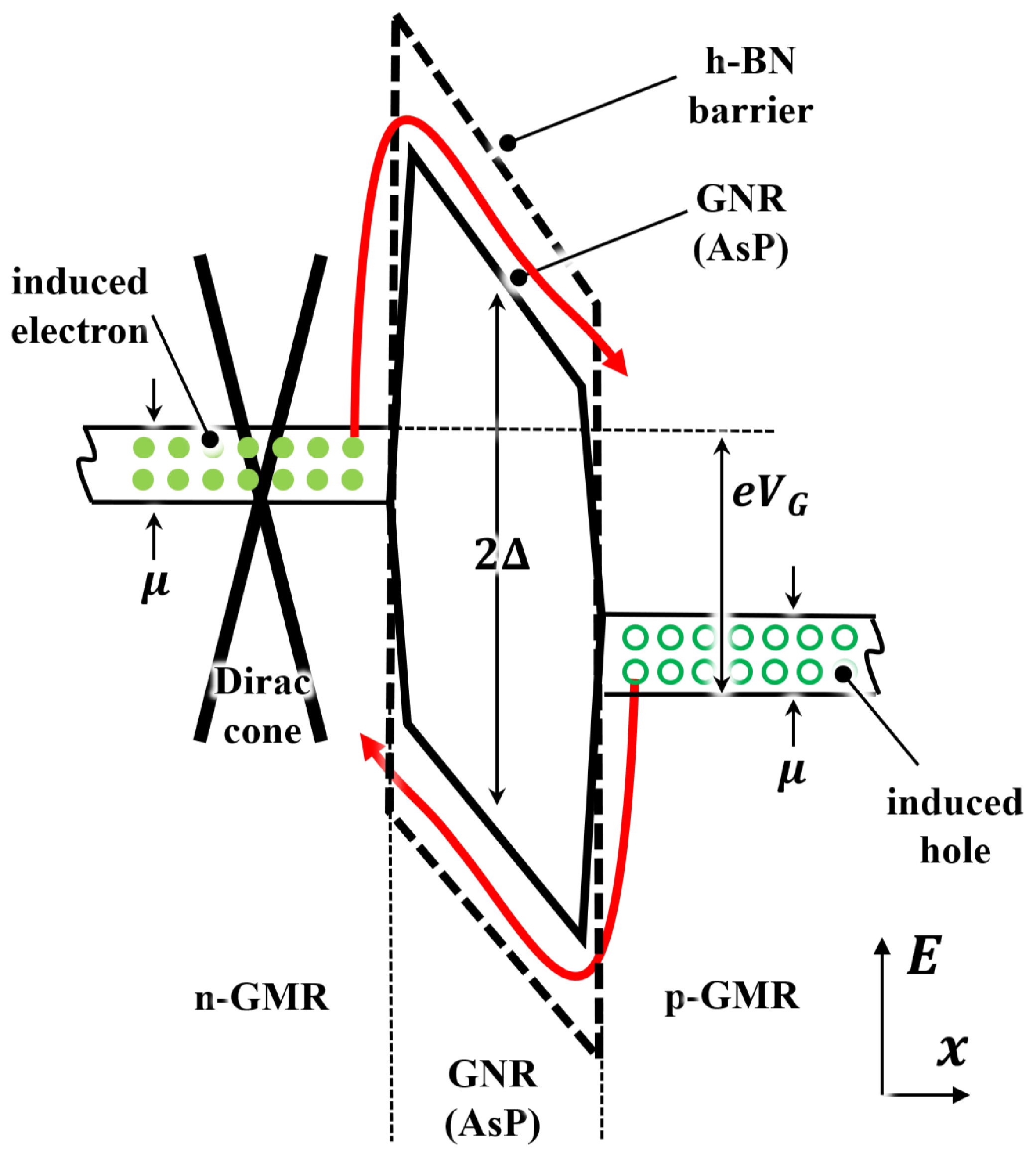}}
\caption{ Band diagram of the  structures under consideration with the trapezoid energy barriers in  the bridges (one period).}
\label{fig2}
\end{figure}

\section{Main assumptions and equations of the model}

Our consideration is based on the following assumptions, which simplify 
the theoretical treatment without compromising 
  model accuracy:\\
(i) The  GNR, CNT, and p-AsP  barriers have the abrupt trapezoidal shape modified by the bias voltage. This leads to
 different device characteristics compared to the devices with  smooth (near parabolic)
 barriers~\cite{24,25};\\
(ii) The inter-GMR current is of the thermionic origin (the tunneling current is disregarded at the structural parameters (in particular, the length of the NBs) and bias voltages under consideration - the pertinent issue is
analyzed in Sec.~IV). In such situations, the thermionic inter-GMR currents are limited by
the carrier activation energy determined by the barrier height $\Delta$, the carrier Fermi energy,  and the effective carrier temperature in GMRs.
The electron and hole Fermi energies and the electron and hole effective temperatures are equal being $\mu_G$ and $T$, respectively;\\
(iii) The injected hot carriers are Maxwellized due to the scattering followed by  energy relaxation;\\ 
(iv) The interaction of the electrons and holes with the optical phonons
in the respective GMRs  is the primary mechanism of the energy relaxation  (OP mechanism) at not too  high carrier temperatures, \cite{26,27,28,29,30,31,32,33,34,35}  accompanied by
the high-energy relaxation mechanisms. For the concreteness, we assume that  
  the supercollision and plasmonic energy relaxation~\cite{36,37,38} (SC-relaxation) plays the role of high-temperature relaxation mechanisms.\\
(v) Carrier  cooling at the contacts to the pads~(Ref.~39, 40, and references therein) and cooling due to the  acoustic phonons~\cite{41,42,43} are disregarded;\\
(vi) The thermal  conductivity of the substrate, GMRs, and  contacting pads are sufficiently high to have  the lattice temperature close to the ambient temperature $T_0$ at the operating voltages.

These assumptions are adequate for the device structures under consideration and do not essentially limit the validity of the obtained results while simplifying the pertinent mathematics.

Since  the GMR/GNR and GMR/CNT structures can be studied within the same mathematical  framework, in the following we refer only to the GMR/GNR and GMR/AsP structures, focusing mainly on the former.

Using the Landauer-Buttiker formula,
accounting for the two-dimensional carrier spectrum in the GMRs 
and the two-dimensional and one-dimensional carrier spectra  in the b-AsP and GNR bridges, respectively, we obtain the following equations for the inter-GMR currents $J_G^{AsP}$ and $J_G^{GNR}$ associated with the thermionic injection above the barrier:

\begin{eqnarray}\label{eq1}
J_{G}^{GNR} \simeq {\overline J}_G^{GNR}
\exp\biggl(\frac{\mu_G - \Delta}{T}\biggr)\biggl[1-\exp\biggl(-{\frac{eV_G}{T}}\biggr)\biggr].
\end{eqnarray} 
\begin{eqnarray}\label{eq2}
J_{G}^{AsP} \simeq {\overline J}_G^{AsP}
\exp\biggl(\frac{\mu_G - \Delta}{T}\biggr)\biggl[1-\exp\biggl(-{\frac{eV_G}{T}}\biggr)\biggr].
\end{eqnarray} 
Here

\begin{eqnarray}\label{eq3}
{\overline J}_G^{GNR} =
\frac{4(2N-1)eT}{\pi\hbar}, \qquad
{\overline J}_G^{AsP} = \frac{8HeT^2}{\pi^2\hbar^2v_W},
\end{eqnarray} 
where $e$ is the carrier charge and $\hbar$ is the Planck constant.

In line with Ref.~24, we present the equation governing the balance of the energy injected into the GMRs and the energy transferred 
by the electron and hole systems to the lattice as follows:

\begin{eqnarray}\label{eq4}
 V_GJ_{G}^{GNR} = 2H_GL_G\Sigma_G\, R, \qquad  V_GJ_{G}^{AsP} = 2H_GL_G\Sigma_G\,R,
\end{eqnarray} 
with $\Sigma_G$ being the carrier surface density in the GMRs.

For 
the carrier energy relaxation determined by the optical phonons and supercollisions, the energy relaxation rate (per one carrier) can be presented in the following  form (see, for example, Refs.~26 and 30): 

\begin{eqnarray}\label{eq5}
R=
\frac{\hbar\omega_0}{\tau_{OP}}
 \biggl[\exp\biggl(-\frac{\hbar\omega_0}{T}\biggr)-\exp\biggl(-\frac{\hbar\omega_0}{T_0}\biggr)\biggr]
 +\frac{1}{\tau_{SC}}\frac{T^3-T_0^3}{T_0^2}.
%
\end{eqnarray} 
Here, $\hbar\omega_0$ is the optical phonon energy in graphene,   $\tau_{OP}$
and $\tau_{SC}$ are  the characteristic times of the optical phonon  spontaneous emission and the relaxation time   
 due to the SC mechanisms, respectively. These times are related to 
the net  energy relaxation time $\tau^{\varepsilon}$ of   
  "warm" carriers ($T\simeq T_0$) at room temperature $T_0$, as
  $1/\tau^{\varepsilon} = 1/\tau^{\varepsilon}_{OP} + 3/\tau_{SC}$ with $1/\tau^{\varepsilon}_{OP} =(\hbar\omega_0/T_0)^2\exp(-\hbar\omega_0/T_0)/\tau_{OP}$ ($\tau^{\varepsilon}_{OP} \gg \tau_{OP}$). The relative contributions of the relaxation mechanisms in question is characterized by parameter $c = \tau_{OP}^{\varepsilon}/\tau_{SC} $.

Due to fairly complex dependences of  the carrier density $\Sigma_G$ and the Fermi energy $\mu_G$  on the temperature $T$ and the voltage $V_G$,  we employ the following interpolating formulas
derived in Appendix A to avoid the
excessive computational problems::

\begin{eqnarray}\label{eq6}
\Sigma_G \simeq \frac{e^2{\overline V}_GV_G}{\pi\hbar^2v_W^2}\biggl(1+ \frac{\pi^2T^2}{3e^2{\overline V}_GV_G}\biggr),
\end{eqnarray}

\begin{eqnarray}\label{eq7}
 \mu_G \simeq\frac{e\sqrt{{\overline V}_GV_G}}{1 + (2T/e\sqrt{{\overline V}_GV_G})},
\end{eqnarray}
Here, 
${\overline V}_G = {\pi\,C_G\hbar^2v_W^2/2}e^3L_G$ is the voltage characterizing the level of degeneracy of the two-dimensional electron and hole systems electrically induced in the pertinent GMRs, $C_G$ is the capacitance of coplanar GMRs per the GMR unit length (determined by the ratio $L_G/L$ and the substrate dielectric constant~\cite{44,45}), $v_W \simeq 10^8$~cm/s is the carrier velocity in GMRs.

Equations~(1) -(5) 
yield the following formulas 
for the voltage and temperature dependences of  ${\overline J}_R^{GNR}$
 and ${\overline J}_R^{AsP}$:

\begin{eqnarray}\label{eq8}
 J_G^{GNR}(T) =  J_G^{AsP} (T) \simeq {\overline J}_R\biggl(1+ \frac{\pi^2T^2}{3e^2{\overline V}_GV_G}\biggr)\nonumber\\
\times 
\biggl[\exp\biggl(\frac{\hbar\omega_0}{T_0}-\frac{\hbar\omega_0}{T}\biggr)-1 +
c\frac{T^3 - T_0^3}{T_0^3}\biggr]
\end{eqnarray}
with 
\begin{eqnarray}\label{eq9}
 J_R =\frac{2H_GC_GT_0^2}{e\hbar\omega_0\tau_{\varepsilon}}.
\end{eqnarray}
Despite the identical temperature dependences of $J_G^{GNR}(T)$ and $J_G^{AsP}(T)$ given by Eqs.~(1) = (3), the resulting voltage dependences can be different due to somewhat different
voltage dependences of the carrier temperature $T$ for GMR/GNR and GMR/AsP structures.\\

\begin{widetext}
In this case, Eqs.~(1) - (9) yield

\begin{eqnarray}\label{eq10}
T\ln\biggl\{\frac{\gamma^{GNR}}{(2N-1)}\biggl(1+ \frac{\pi^2T^2}{3e^2{\overline V}_GV_G}\biggr) \biggl(\frac{T_0}{T}\biggr)
\biggl[\frac{\displaystyle\exp\bigg(\frac{\hbar\omega_0}{T_0}- \frac{\hbar\omega_0}{ T}\biggr)-1 +c\frac{(T^3-T_0^3)}{T_0^3}
}{1-\displaystyle\exp\biggl(-{\frac{eV_G}{T}}\biggr)}\biggr]\biggr\}
+ \Delta = \frac{e\sqrt{{\overline V}_GV_G}}{1 + (2T/e\sqrt{{\overline V}_GV_G})},
\end{eqnarray}

\begin{eqnarray}\label{eq11}
T\ln\biggl\{\gamma^{AsP} \biggl(1+ \frac{\pi^2T^2}{3e^2{\overline V}_GV_G}\biggr)\biggl(\frac{T_0}{T}\biggr)^2
\biggl[\frac{\displaystyle\exp\bigg(\frac{\hbar\omega_0}{T_0}- \frac{\hbar\omega_0}{ T}\biggr)-1+c\frac{(T^3-T_0^3)}{T_0^3}
}{1-\displaystyle\exp\biggl(-{\frac{eV_G}{T}}\biggr)}\biggr]\biggr\}
 + \Delta =\frac{e\sqrt{{\overline V}_GV_G}}{1 + (2T/e\sqrt{{\overline V}_GV_G})},\qquad
\end{eqnarray}
and

\begin{eqnarray}\label{eq12} 
c =\biggl(\frac{\tau_{op}}{\tau_{sc}}\biggr)\biggl(\frac{T_0}{\hbar\omega_0}\biggr)\exp\biggl(\frac{\hbar\omega_0}{T_0}\biggr), \qquad
\gamma^{GNR} =\frac{\pi\,C_GH_GT_0}{2e^2\omega_0\tau_{OP}^{\varepsilon}},\qquad
\gamma^{AsP}  
=\frac{\pi^2C_G\hbar^2v_W}{e^2\hbar\omega_0\tau^{\varepsilon}_{OP}}\frac{H_G}{H}.
\end{eqnarray}

Equations~(10) and (11), corresponding to the metasurfaces with the b-AsP and GNR bridges, respectively, are similar in form but differ in the dependences of $\gamma^{GNR}$ and $\gamma^{AsP}$ on the structural parameters and they exhibit somewhat different temperature dependences.

\end{widetext}

\begin{figure}[th]
\centerline{\includegraphics[width=6.5cm]{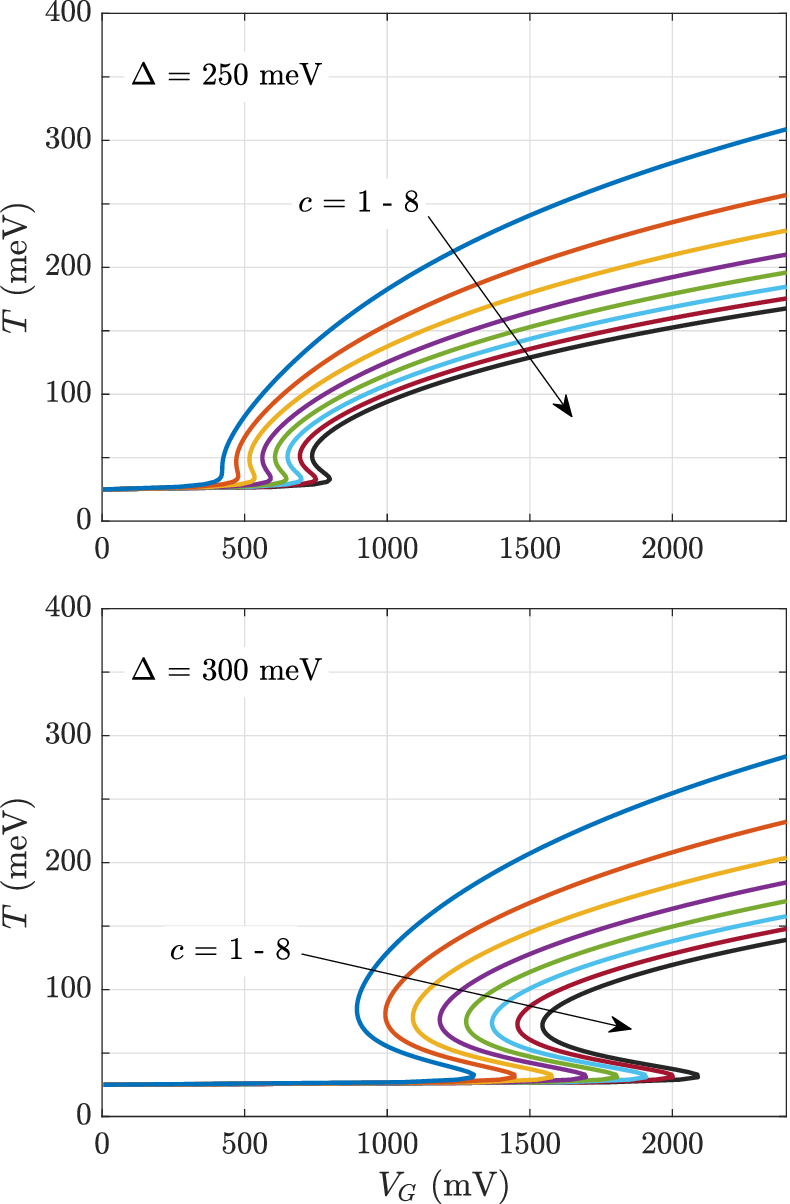}}
\caption{Carrier effective temperature $T$ vs. bias voltage $V_G$ for 
 GMR/GNR structures with $\Delta=250$~meV (upper panel) and $\Delta= 300$~meV (lower panel) at $T_0 = 25$~meV for different values of 
 relaxation ratio $c$.
}
\label{fig3}
\end{figure}

\begin{table}[t]
\centering 
\vspace{2mm}
\begin{tabular}{|r|c|}
\hline
GMR length,\,  $2H_G$& 2~$\mu$m\\
\hline
GNR length,\,  $2L$ &  100~nm\\
\hline
 GMR width, $2L_G$ &  100~nm\\
\hline
GNR  width,\, $2h$ &  (10 - 12)~nm\\
\hline
GNR barrier height, $\Delta$ &  (250 - 300)~meV\\
\hline
Inter-GMR  capacitance, $c_G$& 1.0\\
\hline
Degeneracy voltage, ${\overline V}_G$ & 8 meV\\
\hline 
Structural parameter, $\gamma^{GNR}$ & $8.5\times 10^{-3}$\\ 
\hline  
Number of GNR bridges, $(2N-1)$ & 85 \\ 
\hline
Lattice temperature,\, $T_0$&\, 25~meV \\ 
\hline
Optical phonons energy,  $\hbar\omega_0$& 200~meV \\
\hline
OP energy relaxation   time, $\tau_{OP}^{\varepsilon}$& 20~ps\\
\hline
OP/SC relaxation ratio, $c$ & 1 - 8\\
\hline
\end{tabular}
\caption{\label{table} General GMR/GNR parameters} 
\end{table}

 \section{GMR/GNR  metasurface structure characteristics}

\begin{figure}[t]
\centerline{\includegraphics[width=7.0cm]{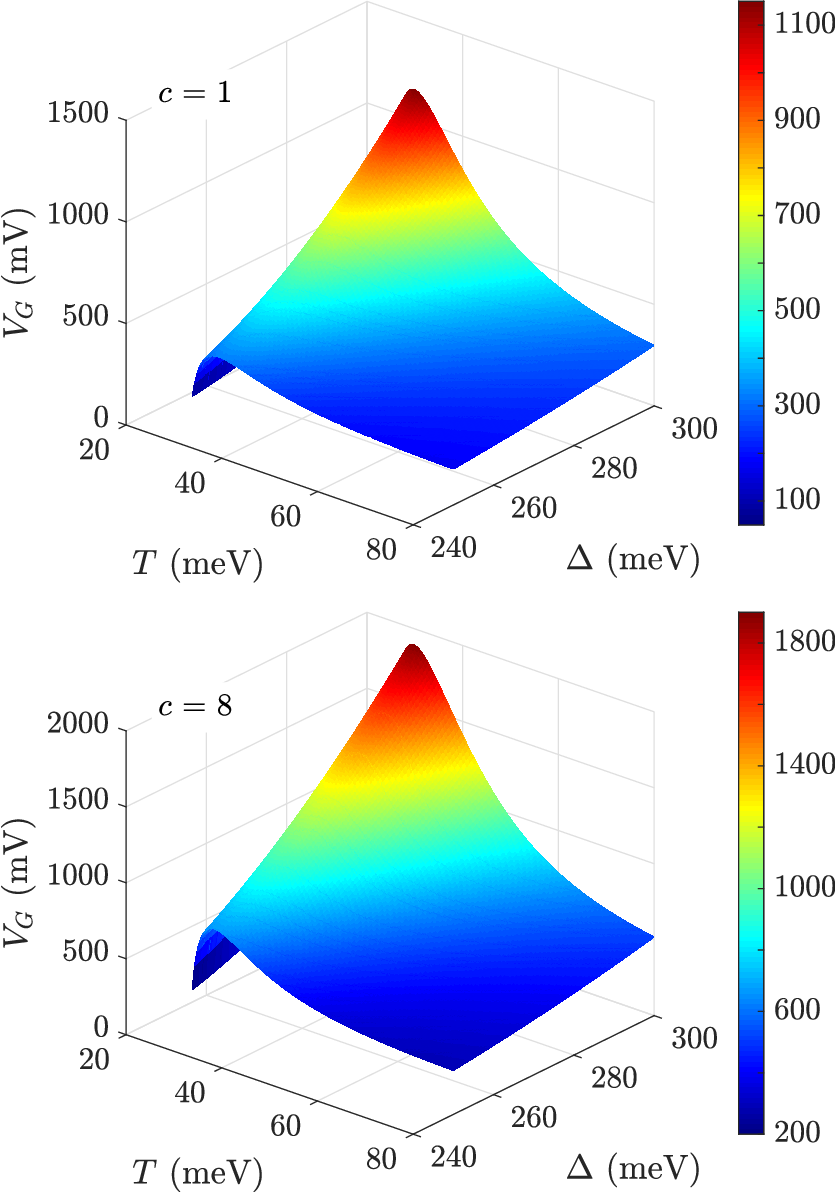}}
\caption{The $V_G - T - \Delta$  relations  for GMR/GNR structures 
with  $c= 1$ (upper panel) and $c = 8$ (lower panel)  at
$T_0 = 25$~ meV.}
\label{fig4}
\end{figure}

\begin{figure}[t]
\centerline{\includegraphics[width=7.0cm]{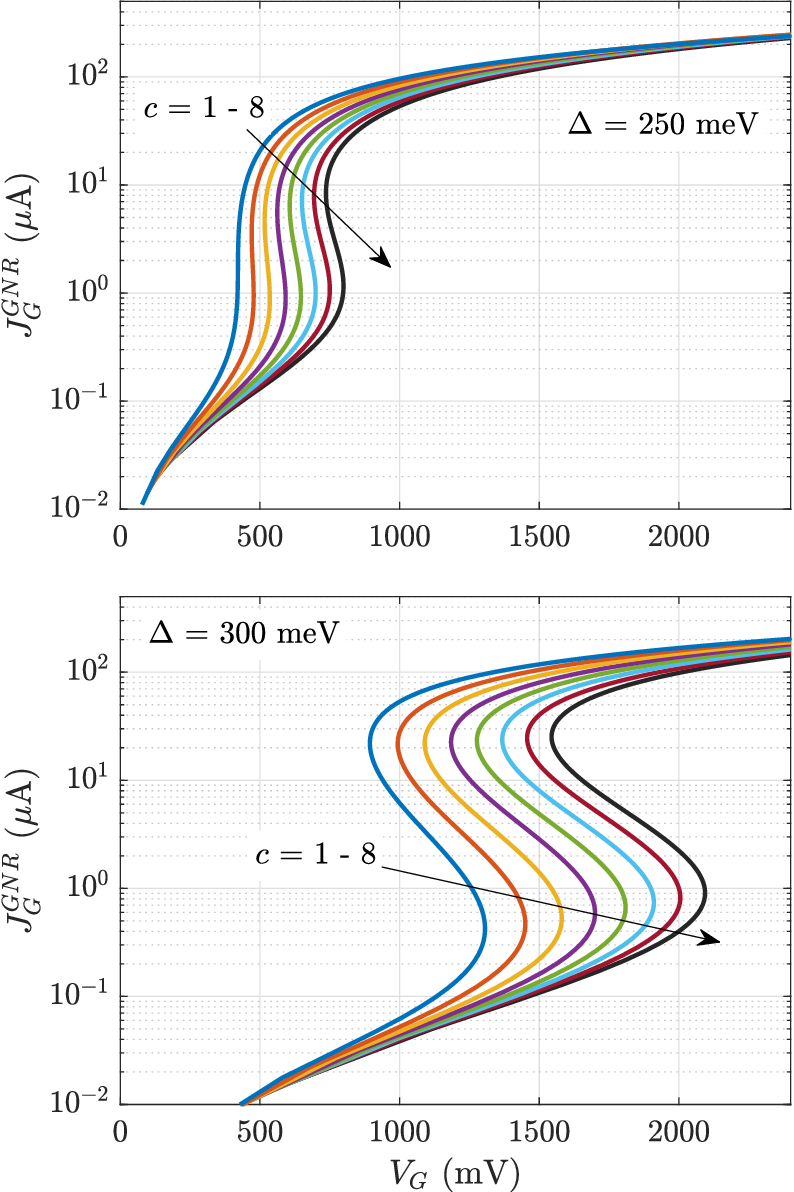}}
\caption{Terminal current (per one GMR pair)  vs. bias voltage $V_G$ corresponding to Fig.~3  for GMR/GNR structures  
 with $\Delta=250$~meV (upper panel) and $\Delta= 300$~meV (lower panel) for different values of 
 relaxation ratio $c$ ($T_0 = 25$~ meV).
}
\label{fig5}
\end{figure}

\begin{figure}[t]
\centerline{\includegraphics[width=7.0cm]{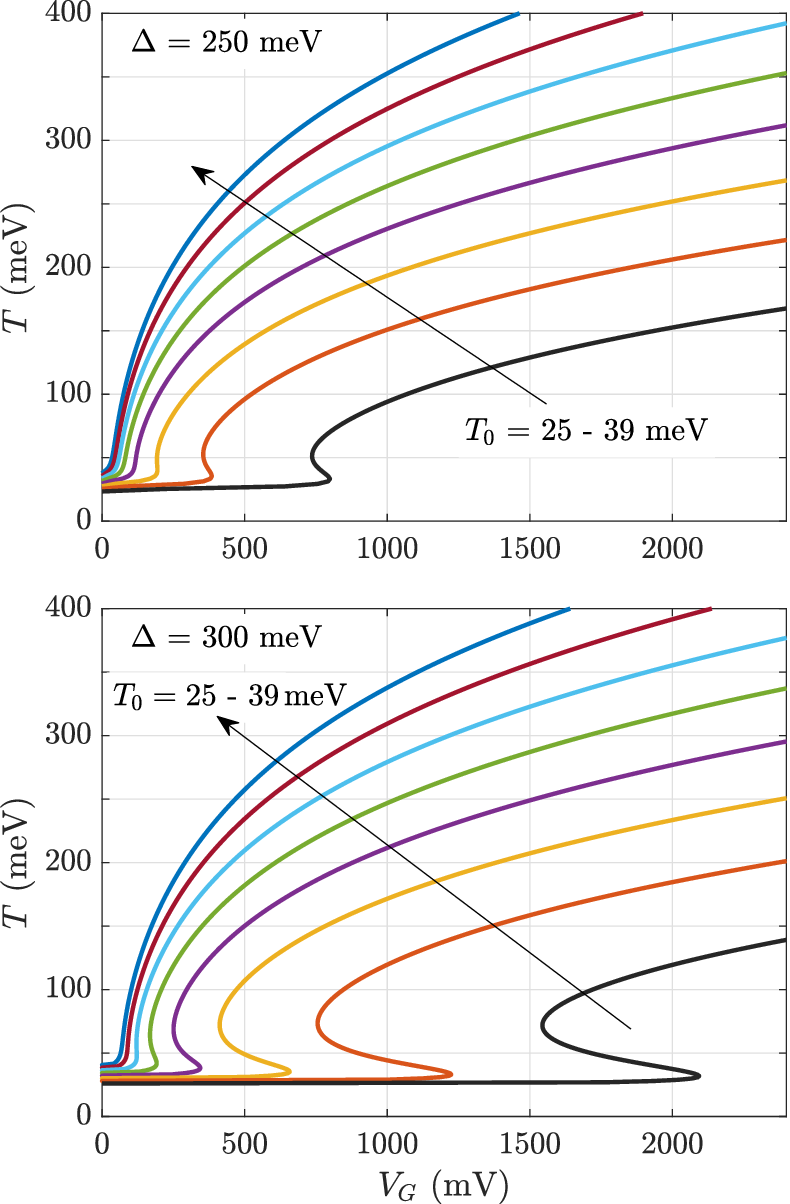}}
\caption{The $T - V_G$  relations 
for GMR/GNR structures with $\Delta = 250$~meV (upper panel),  $\Delta = 300$~meV (lower panel), $c = 8$, and different lattice temperatures  $T_0$ .
}
\label{fig6}
\end{figure}

\begin{figure}[t]
\centerline{\includegraphics[width=7.0cm]{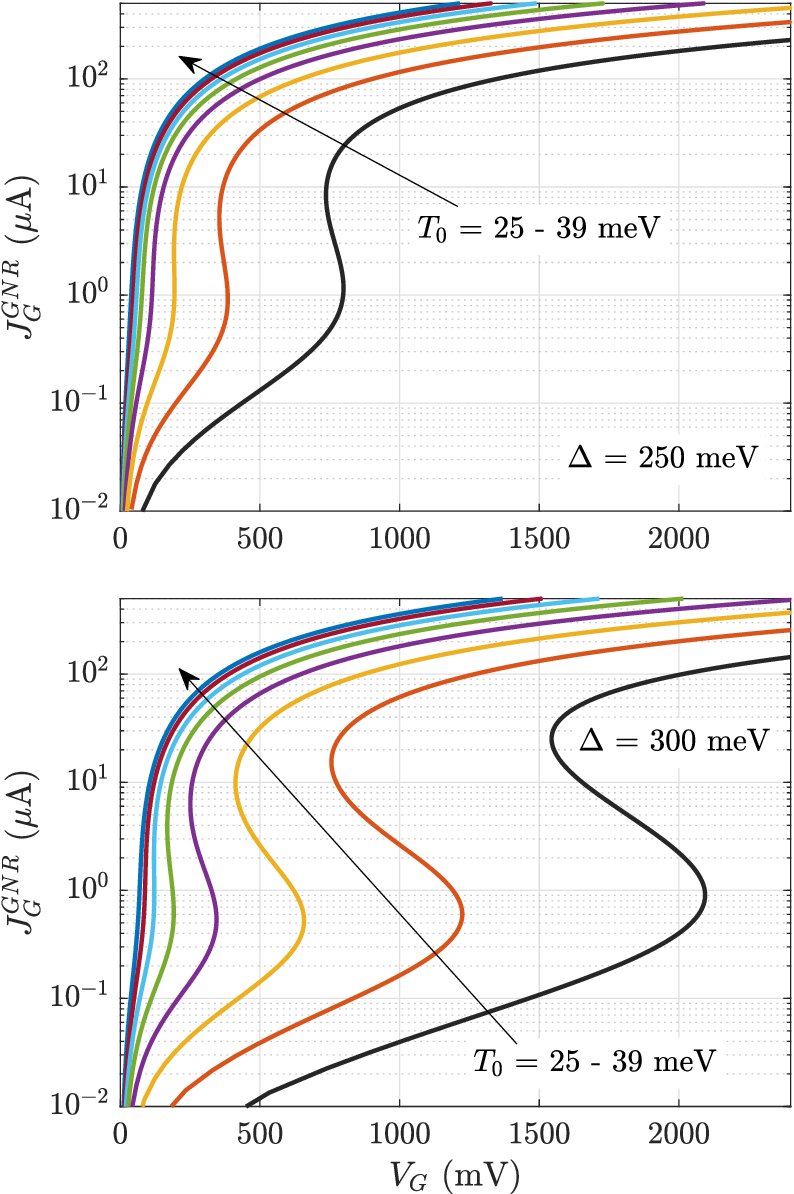}}
\caption{The $J_G^{GNR} - V_G$ relations 
for GMR/GNR structures with $\Delta = 250$~meV (upper panel),   $\Delta = 300$~meV (lower panel),  $c=8$, and different $T_0$.
}
\label{fig7}
\end{figure}

Figure~3  shows the carrier effective temperature $T$ as a function of the bias voltage $V_G$
calculated using Eq. (8)  for the GMR/GNR structures
with $\Delta = 250$~meV and $\Delta = 300$~meV for different values of the OP/SC relaxation ratio $c$ ($c= 1 -8)$. Table I lists  the parameters used  in the calculations. At the chosen values of the GMR length, GNR width $2h$, and  number of GNRs  $2N-1 = 85$, the lateral spacing between the neighboring GNRs is $l  = [H_G -(2N-1)h]/N \simeq 35$~nm. This value is sufficient to prevent the overlap between the carrier states in the neighboring GNRs.
As seen from Fig.~3, the $T - V_G$ relations for the GMR/GNR structures with
 the chosen parameters can be many-valued in a rather wide range of the OP/SC relaxation ratios $c$, exhibiting the branches with
$d T/dV_G > 0$ and $d T/dV_G < 0$ corresponding to low and elevated temperatures
with
$d T/dV_G$ tending to infinity at 
certain voltages $V_G = V_G^{th}$, where $V_G^{th}$ is the threshold voltage dependent on the structural parameters. The threshold voltage corresponds to the threshold temperature $T^{th} = T|_{V_G = V_G^{th}}$ (see below). The abrupt increase in  carrier temperature when $V_G\gtrsim V_G^{th} $
is attributed to hot-carrier thermal breakdown, driven by the positive feedback between
the thermionic current and the carrier temperature.

\begin{figure}[t]
\centerline{\includegraphics[width=7.5cm]{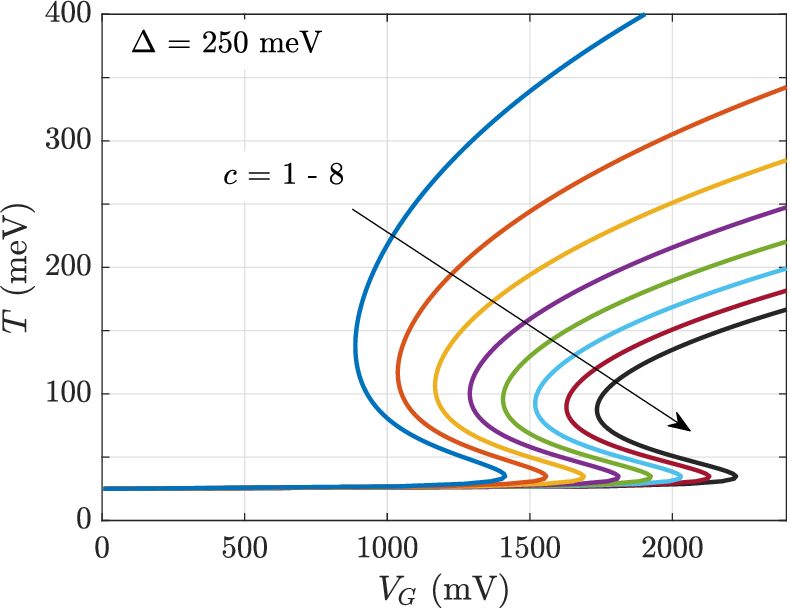}}
\caption{The $ T -V_G$  relations  for GMR/AsP structures with $\Delta = 250$~meV, $H_G/H =1$,
and   different $c$  ($T_0 = 25$~meV).}
\label{fig8}
\end{figure}

\begin{figure}[t]
\centerline{\includegraphics[width=7.0cm]{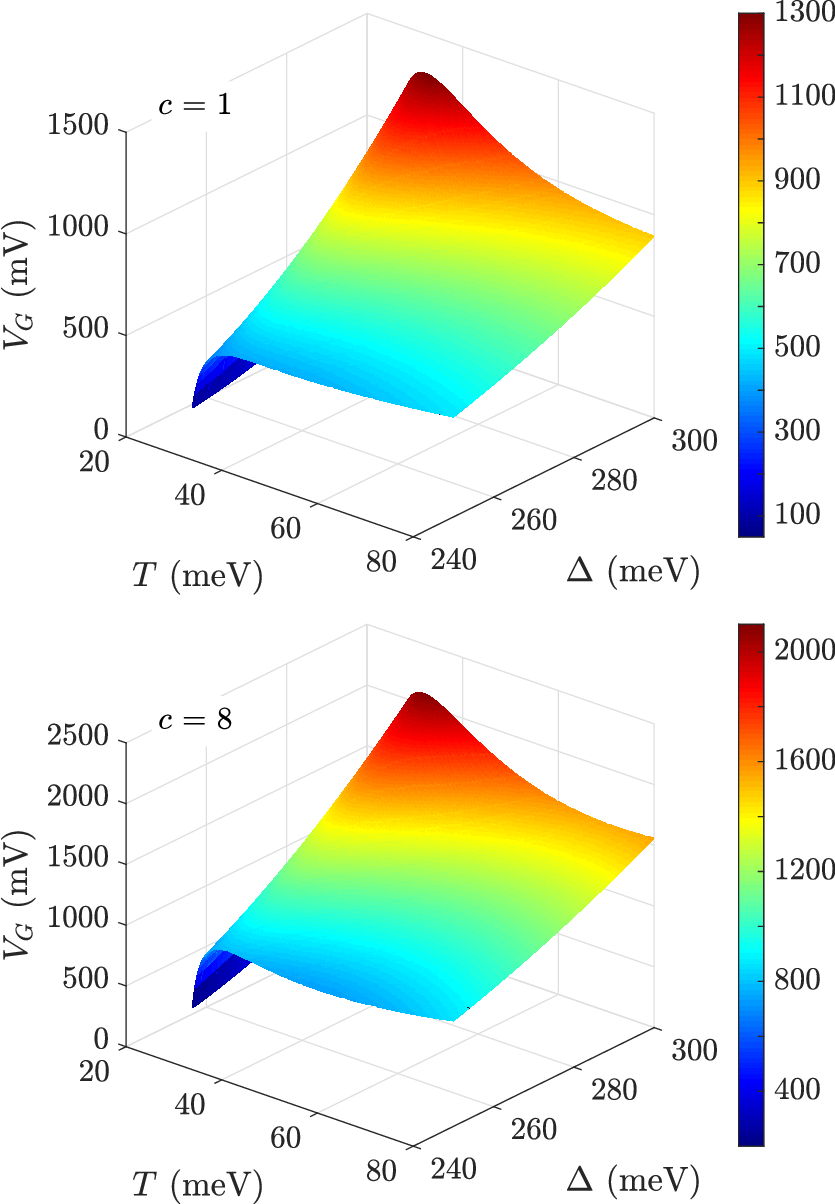}}
\caption{The $V_G - T - \Delta$ relations
for GMR/AsP structure with $H_G/H =1$,  $c= 1$ (upper panel), and $c = 8$ (lower panel) at $T_0 = 25$~meV .}
\label{fig9}
\end{figure}

\begin{figure}[t]
\centerline{\includegraphics[width=7.5cm]{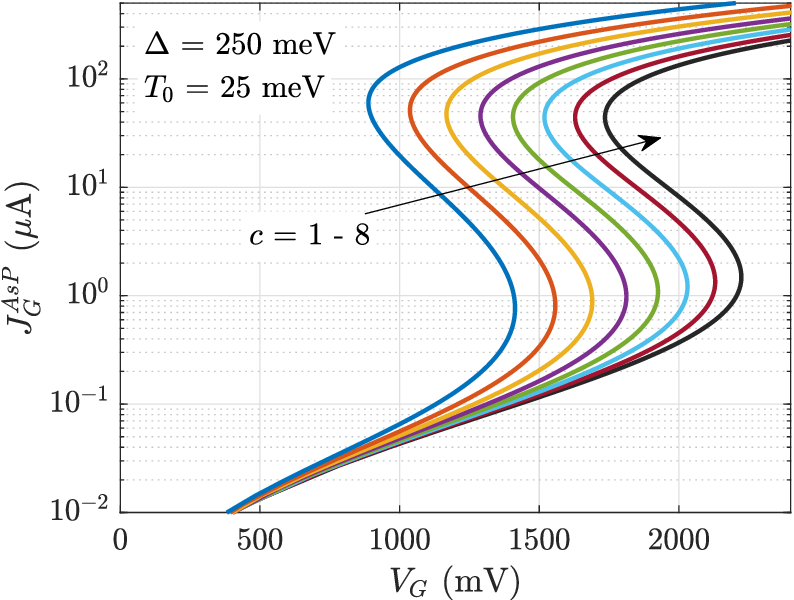}}
\caption{The $ J_G^{AsP} -V_G$  relations  for GMR/AsP structures with $\Delta = 250$~meV, $H_G/H =1$, and different $c$.}
\label{fig10}
\end{figure}
\begin{figure}[th]
\centerline{\includegraphics[width=7.5cm]{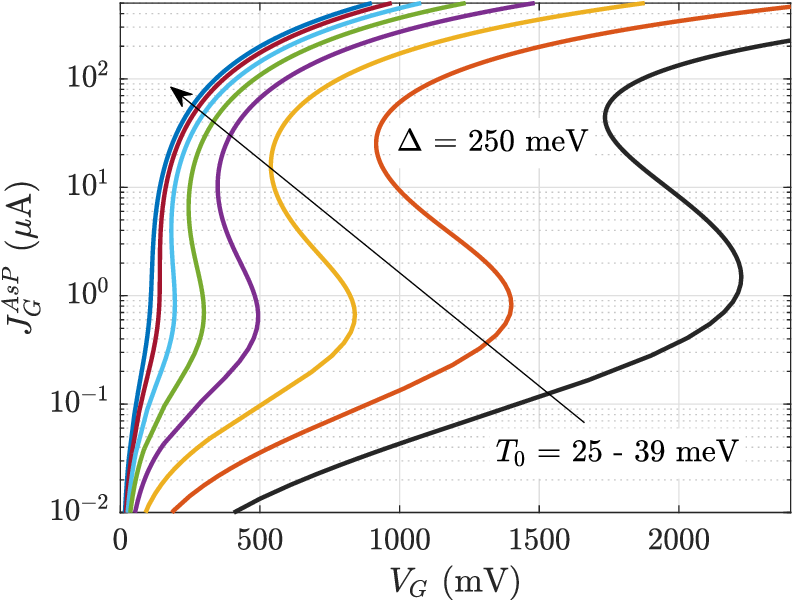}}
\caption{The $J_G^{AsP} - V_G$ relations 
for GMR/AsP structure with $\Delta = 250$~meV,  $H_G/H =1$  and  $c=8$ for different lattice temperatures $T_0$.}
\label{fig11}
\end{figure}

  When $V_G$ exceeds the threshold value $V_G^{th}$, the carrier energy relaxation mechanism associated with optical phonons is unable to limit an increase
in carrier temperature. This increase is stabilized at high temperatures by 
additional scattering mechanisms (namely, by the SC mechanisms), resulting   in the S-shape  $T - V_G$ relations.
Such an  effect is known as  hot-carrier thermal breakdown.
 
As might be expected, the S-shaped  $T - V_G$  characteristics are more pronounced and characterized by larger values of $V_{G}^{th}$ in
the structures with a larger $\Delta$ (compare the dependences for
$\Delta = 250$~meV and  $\Delta = 300$~meV in Fig.~3). The same takes place when  parameter $c$ increases.
An increase in $V_G^{Th}$ is accompanied by rising $T^{th}$ with a relatively moderate increase ($T^{th} \gtrsim T_0 $).
 In contrast, the temperature, $T^{*th}$, at the higher branch at $V_G = V_G^{th}$
decreases with increasing parameter $c$. This is because the higher valies of$c$ correspond to a more substantial contribution from SC energy-relaxation mechamisms (smaller $\tau_{SC}$), which effectively limits the carrier temperature.

Figure~4 shows the $V_G - T - \Delta$ relations calculated for the GMR/GNR structures with  the parameter $c$ characterizing 
the relative contribution of the OP and SC.

Figure 5 shows the voltage dependence of the terminal current $J_G^{GNR}$ calculated using Eq.~(11),
invoking the results of Eq.~(8) solution for the same parameters as for Fig.~3. 
As seen from Figs.~3, 5, 8, and 10,  the current-voltage   characteristics $J_G^{GNR} - V_G$
and  $J_G^{GNR} - V_G$ qualitatively reproduce
the shape of $T - V_G$ relations, in particular, exhibiting the same thermal breakdown voltage values. 
The net terminal current is equal to $MJ_G^{GNR}$, where the number, $M$, of  the GMR pairs can be large.
The shape  of the characteristics  is to grater extend determined by the cumulative parameter $\gamma^{GNR}/(2N-1)$. An increase in the number of the GNRs $(2N-1)$ 
leads to more pronounced S-shape behavior.
 
\begin{table}[b]
\centering 
\vspace{2mm}
\begin{tabular}{|r|c|}
\hline
GMR length,\,  $2H_G$& 4~$\mu$m\\
\hline
AsP bridge  length,\,  $2L$ &  100~nm\\
\hline
 GMR width, $2L_G$ &  100~nm\\
\hline
AsP bridge  width,\, $2H$ &  $~2\mu$m\\
\hline
Asp barrier height, $\Delta$ &  250~meV\\
\hline
Inter-GMR  capacitance, $c_G$& 1.0\\
\hline
Degeneracy voltage, ${\overline V}_G$ & 8 meV\\
\hline 
Structural parameter, $\gamma^{AsP}$ & $0.6745\times 10^{-3}$\\ 
\hline
Lattice temperature,\, $T_0$&\, (25- 39)~meV \\ 
\hline
Optical phonons energy,  $\hbar\omega_0$& 200~meV \\
\hline
OP energy relaxation   time, $\tau_{OP}^{\varepsilon}$& 20~ps\\
\hline
OP/SC relaxation ratio, $c$ & 1 - 8\\
\hline
\end{tabular}
\caption{\label{table} General GMR/AsP parameters} 
\end{table}

\section{GMR/AsP metasurface structures characteristics}

The consideration of the GMR/AsP structures with the same values of the energy barrier heights
and the OP/SC relaxation ratio, based on Eqs.~(2) - (8) results in the $T -V_G$,
$V_G - T - \Delta$, and $J_G^{AsP} - V_G$ characteristics shown in Figs.~8 - 11 calculated using the GMR/AsP parameters listed in Table II. These characteristics
are
qualitatively (but not quantitatively) similar to those in Figs. 3 - 7 for the GMR/GNR.
 The main difference is associated with somewhat different
temperature dependences of the current via the NBs, as follows from Eqs.~(1) - (3). 

\section{Hysteresis}

As seen from the  obtained plots, the $T - V_G$, $J_G^{GMR} - V_G$, and $J_G^{AsP} - V_G$ characteristics exhibit a hysteresis behavior in the wide ranges of the structural parameters. The pertinent hysteresis loops are characterized by the threshold voltages $V_G^{th}$ and $V_G^{*th}$ at which $dT/dV_G$ tends to infinity at the $T - V_G$ characteristics the  lower and upper branches, respectively ($V_G^{*th} < V_G^{th}$). For the device parameters used in the calculations, $V_G^{th}$ varies in the range $(800 - 2100)$~meV. The threshold voltages $V_G^{th}$ and  $V_G^{*th}$ correspond to the respective value of the threshold carrier temperature $T^{Th} \gtrsim T_0$ and $T^{*th} \gg T_0 $.
The current-voltage characteristics,  $J_G^{GMR} - V_G$ and $J_G^{AsP} - V_G$, 
have the threshold currents $J^{th}$
and $J^{*th}$  (with $J^{th} \ll J^{*th} \simeq (100 - 200)~\mu$m)   both corresponding to the same threshold voltage  $V_{G}^{th}$ as the $T - V_G$ characteristics.
This is because

\begin{eqnarray}\label{eq13}
\frac{d J_G^{GNR}}{d V_G} = \frac{d J_G^{GNR }}{dT}\frac{dT}{d V_G} \propto \frac{dT}{d V_G},
\end{eqnarray}

\begin{eqnarray}\label{eq14}
\frac{d J_G^{AsP}}{d V_G} = \frac{d J_G^{Asp}}{dT}\frac{dT}{d V_G} 
\propto \frac{dT}{d V_G},
\end{eqnarray}
so that the voltage derivatives  of the current and the carrier temperature in both the GMR/GNR and GMR/AsP structures 
tend to  infinity simultaneously (i.e., at the same threshold voltage).
Since the states corresponding to the branches of these characteristics 
with $dJ_G/dV_G <0$ are electrically unstable, an increase in $V_G$ beyond $V_G^{th}$ causes the transition from the low temperature and current states to the high temperature and current states.
The reverse transition happens when $V_G$ becomes sufficiently small $V_G < V_G^{*th}$
with $V_G^{*th}$ determined by the high-temperature carrier energy relaxation mechanisms.

 The  hysteresis-type voltage-controlled abrupt transfer of the GMR/GNR  and GMR/AsP metasurface structures   between
the low-temperature/low-current and  the high-temperature/high-current states 
can be used for  fast  switching.  The switching time is determined by optical phonon relaxation mechanism and can be 
 several tens of picoseconds.  Due to high carrier effective temperatures at the upper branch, the hysteresis effect can be used for developing  incandescent terahertz and infrared radiation 
sources. The  GMR/GNR and GMR/AsP operation linked to effective interband radiative transitions,  can offer advantages
 over
other GL-based incandescent radiation emitters.~\cite{46,47,48,48,50,51,52,53,54,55,56}.\\

\begin{table*}[t]
\centering 
\begin{tabular}{|r|c|c|c|c|c|c|c|c|}
\hline 
Device characteristics& $\Delta$, meV  & $V_G^{th}$, mV &  $J^{*th}, \mu$A& 
$C$, kW/cm$^2$K & $P^{*th}/S, \mu$W/cm$^2$ & $\delta T_0$, K&$\delta T_0$, meV\\
\hline
GMR/GNR-A\,&250 & 800&30 &0.5 & 24&  14.4 & 1.24\\
\hline
GMR/GNR-B\,&250 & 800 & 30& 3.0& 24& 2.4 &0.2\\
\hline
GMR/GNR-C\,&300 & 2100& 100&0.5& 210 &42&3.6\\
\hline
GMR/GNR-D\,&300 & 2100 & 100 &3.0 & 210&7&0.6\\
\hline
GMR/AsP -A\,&250 & 2100& 200& 0.5&  420& 84 &7.2\\
\hline
GMR/AsP -B\,&250 &  2100& 200& 3.0& 420&14&1.21\\
\hline
\end{tabular}
\caption{\label{table} Lattice heating ($c = 8$ and  $S = 1\times 10^{-8}$~cm$^{2}$)}. 
\end{table*}

\section{Lattice (substrate) heating} 
The heating of carriers might be accompanied by the lattice heating  associated with, in particular, the optical phonons decay.~\cite{57,58,59,60,61,62}
Due to heat drainage via the h-BN substrate and the terminal contacts,
the  lattice temperature can only slightly exceed the ambient temperature, which is assumed to be equal to $25$~meV ($\simeq 300$~K). Nevertheless, even a moderate lattice heating can affect
the characteristics of the devices under consideration. Figure~6 shows the transformation of the $T -V_G$  and $J_G^{GNR} - V_G$ characteristics associated with the lattice heating. As seen from Fig.~6, 
the S-shape could vanish due to lattice heating.
However, the sharp rise
in  carrier temperature and the current associated with thermal breakdown 
is preserved.

We estimate the deviation of the lattice temperature $\delta T_0$ from the
ambient temperature $T_0$ using the following formula:
$\delta T_0 = P/C  S $,
where $C$ is the thermal conductivity (in units kW/cm$^2$K) accounting for the vertical heat transfer via the h-BN layer and lateral heat transfer,
and $P$ is the Joule power of one
device period with the area $S = 4(H+W)(L_G+ L)$,  where $W$ is the width of the conducting pad.

Table III presents the values  of $\delta T_0$ calculated for the GMR/GNR and GMR/AsP structures
using the parameters from Tables I and II and Figs.~5 and 10. We assumed that $c=8$ and $S = 1\times 10^{-8}$~cm$^2$  ($W = 1~\mu$m). Calculating the Joule power, $P^{th}$, we considered the threshold values of the currents,  $J^{*Th}$, corresponding to the  upper branches of the current-voltage characteristics,  at the respective threshold voltages $V_{G}^{th}$.
The  quantities $\delta T_0$ listed in  Table III are within
the range corresponding to the preservation of the thermal breakdown effect and the S-shaped characteristics.

As seen from Figs.~6 and 10, if  $c \sim 1$, the quantities $V_G^{th}$, $J^{*th}$, and, hence,  the Joule power $P^{th}$ at the threshold voltage 
become smaller. This implies that the lattice heating weakens when $c$ increases.

Thus, with a sufficiently effective heat sink, the variations of the voltage $V_G$  result  in substantial  variations of the carrier temperature $T$, while the lattice temperature remains  close to the ambient temperature $T_0$. In this case,
the  voltage-controlled switching  time of the carrier temperature and the current 
is determined by the carrier energy relaxation (i.e., about several tens of picoseconds). Due to the relatively low inter-GMR capacitance and the high GMR  and contact pads conductances,
the respective recharging RC-delay can be even shorter.

If  the  heat sink is less effective, the temperature of the h-BN
 layer and the GMR lattice (as well as the carrier temperature)  can be much higher than the ambient temperature. As a result, the hysteresis effect in question vanishes (see the curves corresponding to $T_0 \gtrsim 35$~meV in Figs. 6,7, and 11). Even in this case,  a strong sensitivity of the carrier temperature and current to the voltage variation can  be used in applications. However, the  characteristic thermal time constant, which in such a situation is determined by heating and cooling of the h-BN layer, can be much longer. Indeed, 
 for the h-BN layer thickness $t = 100$~nm, the area $S = 1~\mu$m$^2$,  the h-BN heat capacitance  $C_{h-BN} \sim 1000 J$/kgK and BN density  $\rho = 2100$~kg/m$^3$,
we obtain the thermal time constant $\tau_{therm}  \sim 35$~ns.
Hence, if the voltage pulses with the duration $ \sim 100$~ps separated by the intervals of a few $\tau_{therm}$ (about of 100~ns to cool between pulses, one can reduce $\delta T_0$ almost by two orders of magnitude. Hence, in the pulse regime, the S-shaped characteristics can be observed even when the heat sink is far from ideal.
In  the suspended  structures under consideration, the carrier and h-BN layer 
temperatures can be very close been fairly high. In such strictures, the energy balance can be maintained essentially by the radiation processes.
The consideration of this situation is beyond the scope of the present work.

\section{Discussion and Comments}

{\bf Effect of the inter-GMR barriers shape.} 
As shown above, the thermal hot-carrier breakdown effect in the structures with trapezoidal (sharp) inter-GMR barrier reveals if the number, $(2N-1)$, of the inter-GMR
(GNR or CNT) bridges  or if  the width, $2H$, of the AsP bridge 
are  sufficiently large [$(2N-1)$  is about tens, or $H \lesssim H_G$). This is in contrast to the GMR/GNR structures with smooth (near parabolic barriers) considered in Ref.~24, in which the thermal hot-carrier breakdown is possible at moderate  $(2N-1$) (a few bridges). Apart from this,
the effect is revealed at markedly higher bias voltages

{\bf Inter-GMR tunneling.}  
Above, we disregarded the inter-GMR tunneling via the GNR and AsP bridges.
The rough estimate (see Appendix B) with the logarithmic accuracy shows that the latter  is justified if

\begin{eqnarray}\label{eq15}
\exp\biggl(-\frac{\Delta}{T}\biggr) \gg  \exp\biggl(-\frac{V_{tunn}^{GNR}}{V_G}\biggr),\quad\exp\biggl(\frac{V_{tunn}^{AsP}}{V_G}\biggr).
\end{eqnarray}
Here, $V_{tunn}^{GNR}= (2L\Delta^2/e\hbar\,v_W)$ and 
$V^{AsP}_{tunn} = (8L\sqrt{2m_{xx}}\Delta^{3/2}/3e\hbar)$ are the respective tunneling voltages with $m_{xx}$ being the carrier effective mass in the AsP layer in  the tunneling direction.
Inequalities~(15) correspond to

\begin{eqnarray}\label{eq16}
T \gtrsim \quad T_{tunn}^{GNR}, \quad T_{tunn}^{AsP}
\end{eqnarray}
with

\begin{eqnarray}\label{eq17}
T_{tunn}^{GNR} = \frac{\hbar\,v_W}{2L}\frac{eV_G}{\Delta}, \quad T_{tunn}^{AsP }=\frac{\hbar}{2L}\sqrt{\frac{9\Delta}{32m_{xx}}} \frac{eV_G}{\Delta}.
\end{eqnarray}

For the GMR/GNR structure with $2L = 100$~nm, $\Delta = (250 -300)$~meV  (as assumed above),  $m_{xx} = 5\times 10^{-29}$~g,
in the  voltages close to the threshold values   $V_G = (800 - 2000)$~meV (see Figs. 3 and  5),
we obtain $T_{tunn}^{GNR} \simeq 20 -40$~meV. 
For the GMR/AsP structure with $2L = 100$~nm, $\Delta = 250$~meV,  $m_{xx} = 5\times 10^{-29}$~g, for $V_G = (200 - 2250)$~meV (see Figs.~8 and 10),
we obtain $T_{tunn}^{AsP} \simeq (3 -30)$~meV. 
These estimates confirm that in the GMR/GNR and GMR/AsP structures with relatively long inter-GMR bridges, $2L = 100$~nm, the carrier tunneling does not affect  the  lower branches of the obtained S-shaped characteristics and, particularly, of the upper branch. 
In the structures with much smaller inter-GMR spacing, the tunneling can both quantitatively and qualitatively influence the S-shaped characteristics. 
 
{\bf Role of interface optical phonons.} The model used above can be generalized by including  the carrier energy relaxation on surface optical phonons. However, since the energy of the latter $\hbar\omega_S < \hbar\omega_0$ and the pertinent energy relaxation time exceeds $\tau^{\varepsilon}$,
at $\Delta > \hbar\omega_0 > \hbar\omega_S$, such an extension of the model does not qualitatively modify the obtained results (analogously to Ref.~62) and does not change quantitatively 
if consider $\tau^{\varepsilon}$ characterizing both types of optical phonons.

\section{Conclusion}
We  analyzed  GMR/GNR (GMR/CNT) and GMR/AsP metasurfaces, consisting of the coplanar interdigital GMR arrays arranged in pairs connected by NBs, and analyzed the dependences of the carrier effective temperature and the terminal current on the bias voltage.
The obtained voltage characteristics can be many-valued (of S-type).
This is associated with the positive feedback between the inter-GMR thermionic currents
through the NBs  and carrier heating, leading to hot-carrier thermal breakdown.  The heating of the substrate can modify the  voltage characteristics, preserving, however, their shape over a relatively wide range of the heat sink parameters.The features of the obtained device characteristics can be used in fast voltage-controlled current switches and fast-modulated  incandescent
sources of  terahertz and infrared radiation.

\section*{ACKNOWLEDGMENTS}
The work 
at Tohoku University  and
University of Aizu was supported by the Japan Society for
the Promotion of Science (KAKENHI Grants Nos.25K01281  and 25K22093),
 The Telecommunications Advancement Foundation, SCAT, JST ALCA-Next, and NEDO, Japan.
 The work at Rensselaer Polytechnic Institute
  was supported
by AFOSR (Contract No. FA9550-19-1-0355).
The work at ENSEMBLE3 Ltd. was supported under the International Research Agenda program of the Foundation for Polish Science 
(FENG.02.01-IP.05-0044/24), the European Union through the Horizon 2020 Teaming for Excellence program (GA No. 857543), and the Poland Minister of Science and Higher Education Center of Excellence project under the Horizon 2020 program (No. MEiN/2023/DIR/3797).\\

\section*{AUTHOR DECLARATIONS}
{\bf Conflict of Interest}\\
The authors have no conflicts to disclose.

\section*{ DATA AVAILABILITY}
All data that support the findings of this study are available within the article.\\

 \begin{widetext}  
\section*{Appendix A. Interpolating formulas for the carrier density and Fermi energy}
 \setcounter{equation}{0}
\renewcommand{\theequation} {A\arabic{equation}}

The surface density, $\Sigma_G$, of both types of carriers in the p-GMRs
\begin{eqnarray}\label{eqA1}
\Sigma_G = \frac{2T^2}{\pi\hbar^2v_W^2}\int_0^{\infty}\frac{d\varepsilon\varepsilon}{[\exp(\varepsilon - \mu_G/T)+1]} +\frac{2T^2}{\pi\hbar^2v_W^2}\int_0^{\infty}\frac{d\varepsilon\varepsilon}{[\exp(\varepsilon + \mu_G/T)+1]},
\end{eqnarray}
with $$
\Sigma_G \simeq \frac{\pi}{3}\biggl(\frac{T}{\hbar\,v_W}\biggr)^2
$$
if $\mu_G \ll T$, and

 $$
\Sigma_G = \frac{\mu^2}{\pi \hbar^2v_W^2} \simeq\frac{e^2{\overline V}_GV_G}{\pi \hbar^2v_W^2}
$$
if $\mu_G \gg T$.\\

On the other hand,

\begin{eqnarray}\label{eqA2}
 \frac{2T^2}{\pi\hbar^2v_W^2}\int_0^{\infty}\frac{d\varepsilon\varepsilon}{[\exp(\varepsilon - \mu_G/T)+1]} -\frac{2T^2}{\pi\hbar^2v_W^2}\int_0^{\infty}\frac{d\varepsilon\varepsilon}
 {[\exp(\varepsilon + \mu_G/T)+1]} = \frac{C_GV_G}{2eL_G}. 
\end{eqnarray}

\begin{eqnarray}\label{eqA3}
 \Sigma_G= \frac{C_GV_G}{2eL_G}+ \frac{4T^2}{\pi\hbar^2v_W^2}\int_0^{\infty}\frac{d\varepsilon\varepsilon}{[\exp(\varepsilon - \mu_G/T)+1]} \simeq+frac{C_GV_G}{2eL_G}+\frac{\pi}{3}\biggl(\frac{T}{\hbar\,v_W}\biggr)^2 \nonumber\\
= \frac{C_GV_G}{2eL_G}\biggl[1+      \frac{2\pi\,eL_G}{3C_GV_G}\biggl(\frac{T}{\hbar\,v_W}\biggr)^2  \biggr]=\frac{e^2{\overline V}_GV_G}{\pi\hbar^2v_W^2}\biggl(1+ \frac{\pi^2T^2}{3e^2{\overline V}_GV_G}\biggr)
\end{eqnarray}

$$
\mu_G \simeq \frac{\pi\hbar^2v_W^2C_GV_G}{4eL_GT} = \frac{e^2{\overline V}_GV_G}{2T}\propto \frac{1}{T}
$$
if $\mu_G \ll T$, and
$$
\mu_G \simeq e\sqrt{{\overline V}_GV_G}
$$
if  $\mu_G \gg T$, where
${\overline V}_G = (\pi\,C_G\hbar^2v_W^2/2e^3L_G)$

Accounting for this, we use the following interpolating formulas for
the dependences of $\Sigma_G$ and $\mu_G$

\begin{eqnarray}\label{eqA4}
\Sigma_G \simeq \frac{e^2{\overline V}_GV_G}{\pi\hbar^2v_W^2}\biggl(1+ \frac{\pi^2T^2}{3e^2{\overline V}_GV_G}\biggr)
,\qquad
 \mu_G \simeq\frac{e\sqrt{{\overline V}_GV_G}}{1 + (2T/e\sqrt{{\overline V}_GV_G})},
\end{eqnarray}

\section*{Appendix B. Lateral tunneling current  through GNR and AsP bridges }
 \setcounter{equation}{0}
\renewcommand{\theequation} {B\arabic{equation}}

If $ T \ll  \mu \simeq e\sqrt{{\overline V}_GV_G} \ll eV_g \sim \Delta$, the tunneling transparency of the GNR is approximately given by~\cite{63}

\begin{eqnarray}\label{eqB1}
T^{GNR}(\varepsilon) \simeq\exp\biggl[ - \frac{V_{Tunn}}{(V_G -2\sqrt{{\overline V}_GV_G})}\biggr],
\end{eqnarray}
which  yields

\begin{eqnarray}\label{eqB2}
J _{tunn}^{GNR} \simeq (2N-1)\frac{4e^2\sqrt{{\overline V}_G(V_G - 2\sqrt{{\overline V}_GV_G})}}{\pi\hbar}\exp\biggl[ - \frac{V_{tunn}^{GNR}}{(V_G - 2\sqrt{{\overline V}_GV_G})}\biggr]
\simeq (2N-1)\frac{4e^2\sqrt{{\overline V}_GV_G}}{\pi\hbar}\exp\biggl( - \frac{V_{Tunn}}{V_G}\biggr). 
\end{eqnarray}

The tunneling transparency of the AsP bridge for the carriers with the energy $\varepsilon$
and momentum $p_z$ is equal to

\begin{eqnarray}\label{eqB3}
T(\varepsilon, p_z)^{AsP} = \exp\biggl[- \frac{2}{\hbar}\int_0^{x_0} dx\sqrt{2m_{xx}\biggl(\Delta - \varepsilon -  \frac{p_z^2}{2m_{zz}} -\frac{e\,(V_G-2\sqrt{{\overline V}_GV_G})}{2L}x\biggr)}
\biggr]
\simeq \exp\biggl[-\frac{8L\sqrt{2m_{xx}}(\Delta -\varepsilon)^{3/2}}{3e\hbar\,(V_G-2\sqrt{{\overline V}_GV_G})}\biggr],\qquad
\end{eqnarray}
where $x_0 = 2L\Delta/e(V_G - 2\sqrt{{\overline V}_GV_G})$, $m_{xx}$ and $m_{zz}$ (due to a strong anisotrophy 
of the b-AsP band structure, in the AsP bridges based on this material $m_{xx} \ll m_{zz}$)  are the pertinent  effective masses in the inter-GMR and transverse directions.. We have neglected the dependence of the tunneling transparency on the lateral momentum $p_z$ because of relatively large $m_{zz}$
(this dependence is  characterized by a small parameter $m_{xx}/m_{zz}$, where $m$ is the fictitious carrier  mass in GMRs). 

Accounting Eq.~(B3),
for the tunneling current through the AsP bridge at $\mu \gg T$ we obtain

\begin{eqnarray}\label{eqB4}
 J^{AsP}_{tunn} \simeq \frac{Hv_Wc_GV_G}{2\pi\,L_G} 
\exp\biggl[-\frac{V_{AsP}^{tunn}}{(V_G-2\sqrt{{\overline V}_GV_G})}
\biggl(1 -\frac{e\sqrt{{\overline V}_G\,V_G}}{\Delta}\biggr)^{3/2}\biggr]
\simeq \frac{Hv_Wc_GV_G}{2\pi\,L_G} 
\exp\biggl(-\frac{V_{AsP}^{tunn}}{V_G}.
\biggr).
\end{eqnarray}

In the above formulas,

\begin{eqnarray}\label{eqB5}
{\overline V}_G =\frac{\pi\,c_G\hbar^2v_W^2}{2e^3L_G}, \quad V_{Tunn}^{GNR} = \frac{\pi\Delta^2L}{\hbar\,v_We}. \quad
V^{tunn}_{AsP} = \frac{8L\sqrt{2m_{xx}}\Delta^{3/2}}{3e\hbar}
\end{eqnarray}
are
the characteristic  voltages. 
\end{widetext}


\end{document}